\documentclass[conference]{IEEEtran}
\IEEEoverridecommandlockouts
\usepackage{cite}
\usepackage{amsmath,amssymb,amsfonts}
\usepackage{algorithmic}
\usepackage{graphicx}
\usepackage{textcomp}
\usepackage{xcolor}

\usepackage{hyperref}
\usepackage{multirow}
\usepackage{subcaption}
\usepackage{booktabs}
\usepackage{enumitem}

\def\BibTeX{{\rm B\kern-.05em{\sc i\kern-.025em b}\kern-.08em
    T\kern-.1667em\lower.7ex\hbox{E}\kern-.125emX}}
\begin{document}

\title{BitROM: Weight Reload-Free CiROM Architecture Towards Billion-Parameter 1.58-bit LLM Inference}

\author{
Wenlun Zhang\textsuperscript{1}, Xinyu Li\textsuperscript{1,2}, Shimpei Ando\textsuperscript{1}, and Kentaro Yoshioka\textsuperscript{1} \\
\textsuperscript{1}Keio University \quad \textsuperscript{2}Nanjing University \\
{\tt\small \{wenlun\_zhang, shimpeiando, kyoshioka47\}@keio.jp} \quad {\tt\small xinyuli@smail.nju.edu.cn}
}

\maketitle

\begin{abstract}
Compute-in-Read-Only-Memory (CiROM) accelerators offer outstanding energy efficiency for CNNs by eliminating runtime weight updates. However, their scalability to Large Language Models (LLMs) is fundamentally constrained by their vast parameter sizes. Notably, LLaMA-7B---the smallest model in LLaMA series---demands more than 1,000 cm\textsuperscript{2} of silicon area even in advanced CMOS nodes. This paper presents BitROM, the first CiROM-based accelerator that overcomes this limitation through co-design with BitNet's 1.58-bit quantization model, enabling practical and efficient LLM inference at the edge. BitROM introduces three key innovations: 1) a novel Bidirectional ROM Array that stores two ternary weights per transistor; 2) a Tri-Mode Local Accumulator optimized for ternary-weight computations; and 3) an integrated Decode-Refresh (DR) eDRAM that supports on-die KV-cache management, significantly reducing external memory access during decoding. In addition, BitROM integrates LoRA-based adapters to enable efficient transfer learning across various downstream tasks. Evaluated in 65nm CMOS, BitROM achieves 20.8 TOPS/W and a bit density of 4,967 kB/mm\textsuperscript{2}---offering a 10$\times$ improvement in area efficiency over prior digital CiROM designs. Moreover, the DR eDRAM contributes to a 43.6\% reduction in external DRAM access, further enhancing deployment efficiency for LLMs in edge applications. \textbf{Code is available at} \href{https://github.com/Wenlun-Zhang/BitROM}{\textit{https://github.com/Wenlun-Zhang/BitROM}}
\end{abstract}


\section{Introduction}

Deep Neural Networks (DNNs) have achieved remarkable success across numerous artificial intelligence-driven applications. However, the rapid increase in both model complexity and parameter counts has posed significant challenges for deploying these powerful networks on resource-constrained edge devices. Traditional hardware architectures frequently transfer weights between processing cores and external DRAM, resulting in a pronounced memory bottleneck and substantial energy overhead. To address this bottleneck, Compute-in-Memory (CiM) has emerged as a promising solution, enabling computation directly within memory cells to substantially reduce data movement and improve energy efficiency.

\begin{figure}[htbp]
    \centering
    \begin{subfigure}{0.47\linewidth}
        \includegraphics[width=\linewidth]{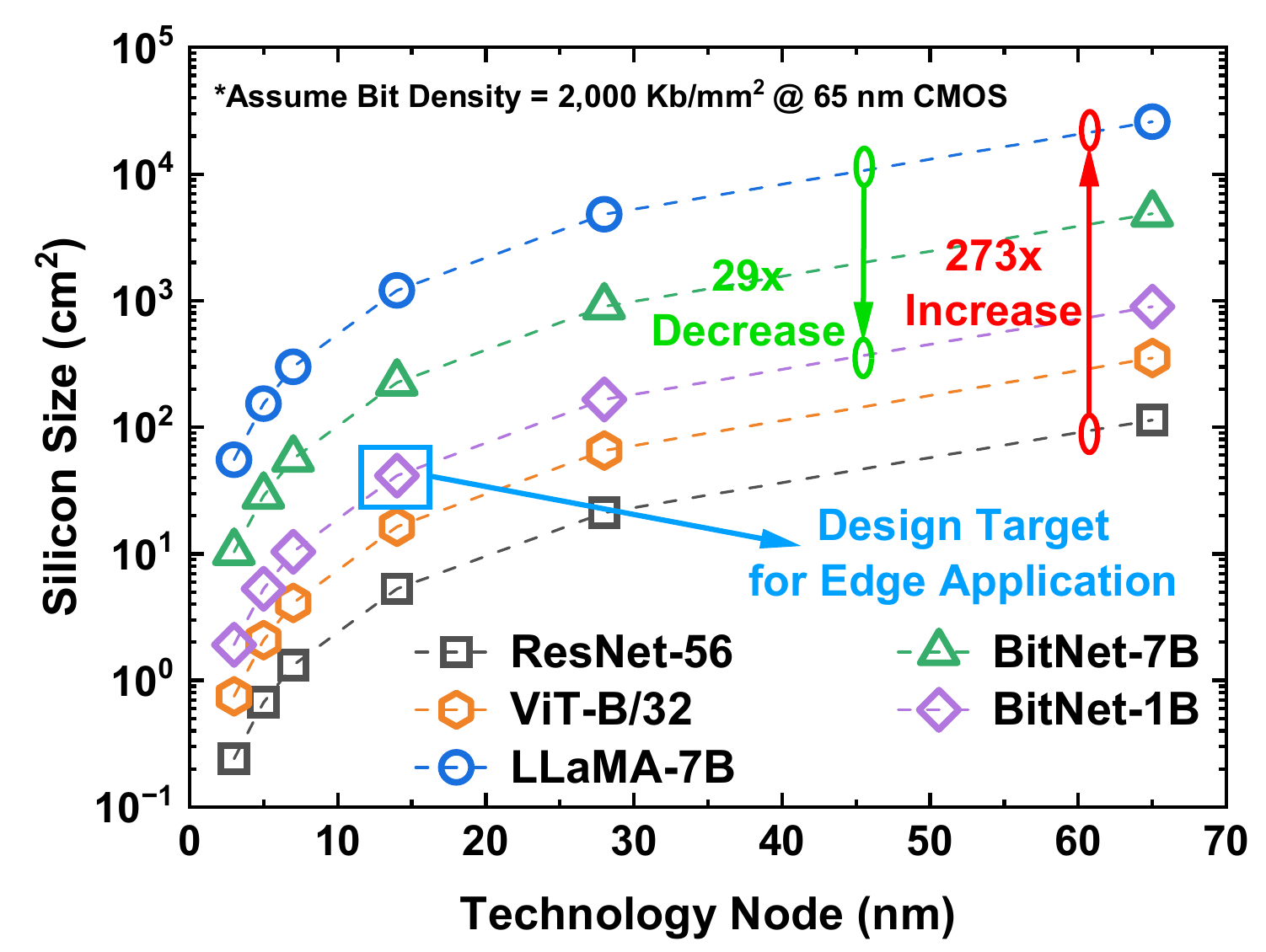}
        \caption{}
        \label{}
    \end{subfigure}
    \hfill
    \begin{subfigure}{0.47\linewidth}
        \includegraphics[width=\linewidth]{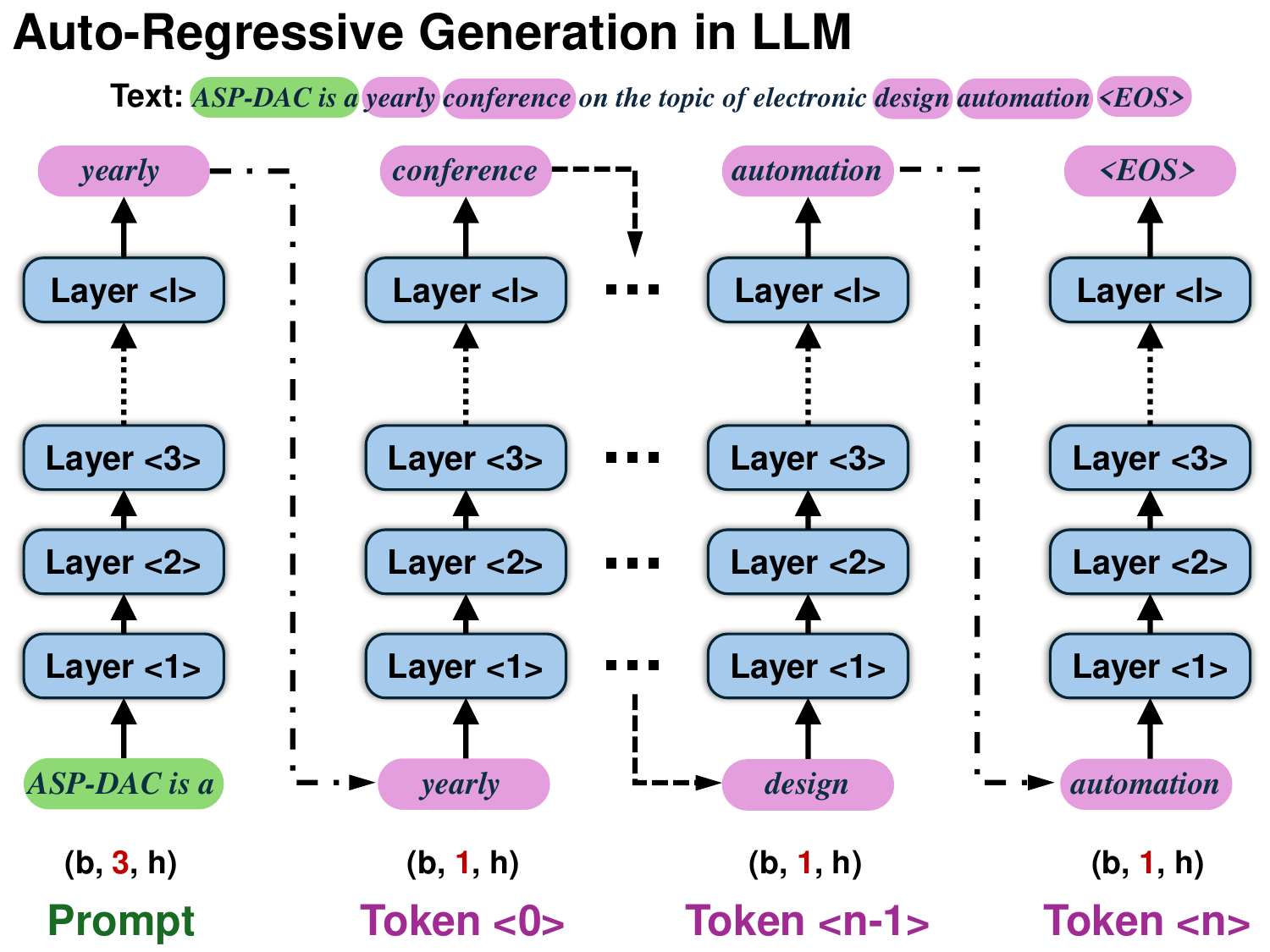}
        \caption{}
        \label{}
    \end{subfigure}
    \caption{(a) Silicon area estimation of CiROM architectures for different model sizes and fabrication nodes. (b) Token-wise auto-regressive generation during LLM inference.}
    \label{Fig_Background}
\end{figure}

Recent CiM accelerators have steadily achieved higher energy efficiency in multiply-and-accumulate (MAC) computations. However, most of these architectures still rely on frequent weight updates within the memory arrays, which somewhat contradicts the fundamental concept of CiM and leads to noticeable system-level efficiency degradation. In pursuit of eliminating the need for weight movement between processing units and external DRAM, Compute-in-Read-Only-Memory (CiROM) has attracted growing interest due to its remarkably high cell density, allowing each cell to be implemented with merely one transistor. Recent CiROM-based studies have demonstrated the feasibility of mapping all ResNet-56 weights within a small silicon footprint~\cite{DCiROM}. However, a fundamental scalability limitation prevents such architectures from supporting Large Language Models (LLMs). For example, even the smallest model in LLaMA series (LLaMA-7B~\cite{Llama2}) requires approximately 273$\times$ more silicon area than ResNet-56, exceeding 1,000 cm\textsuperscript{2} in 14nm CMOS, as illustrated in Fig.~\ref{Fig_Background}(a). Such excessive area requirements make edge integration technically impractical.

Encouragingly, BitNet's ternary quantization offers a promising pathway to close this scalability gap. By reducing LLM weights to approximately 1.58-bit precision, BitNet enables the development of compact yet high-performing models with parameter counts as low as 1B. Notably, the BitNet-1B model reduces the required silicon area to merely tens of cm\textsuperscript{2} in 14nm CMOS---a scale that serves as the design target of this work. Nonetheless, these advancements necessitate a fundamental rethinking of CiROM architectures to address three critical challenges associated with LLM inference:

\begin{itemize}[leftmargin=*]
    \item First, LLMs are highly sensitive to computational errors~\cite{ASiM}, rendering existing area-efficient analog CiROM designs unsuitable for BitNet LLMs~\cite{QLC_CiROM,Hybrid_SRAM_ROM}. Even when weights are quantized to ternary levels, the challenge of accommodating the vast number of parameters in LLMs within a compact CiROM chip remains significant. Moreover, directly deploying ternary weights into existing CiROM macros would result in unnecessary resource and energy overhead, as these circuits are not optimized for ternary-weight models.
    \item Second, unlike conventional DNNs, LLMs exhibit a fundamentally different auto-regressive generation pattern, as illustrated in Fig.~\ref{Fig_Background}(b). During inference, the decode phase leads to progressive expansion of the KV-cache with increasing sequence length, as depicted in Fig.~\ref{Fig_Decode-Refresh_eDRAM}(a). This leads to a growing demand for memory bandwidth to fetch data from external memory, ultimately degrading the system-level efficiency of CiROM accelerators.
    \item Third, CiROM generally lacks flexibility across tasks because weight parameters are fused after fabrication. While some domain-transfer approaches have been proposed for image classification~\cite{YOLoC,Hidden-ROM}, the hardware adaptation of CiROM for domain transfer in LLMs, particularly BitNet models, remains largely unexplored.
\end{itemize}

To address these challenges, we propose BitROM, the first CiROM-based accelerator co-designed with BitNet LLMs. Our key insight is that efficient LLM acceleration necessitates joint optimization across three fronts: ternary computation, auto-regressive memory access patterns, and adaptability to downstream tasks---features not supported by existing CiROM designs. The main contributions of this paper are as follows:

\begin{itemize}[leftmargin=*]
    \item We designed the BitROM accelerator to align with the properties of BitNet LLMs, enabling full-model mapping onto CiROM chips and completely eliminating runtime weight reloading, thereby achieving efficient LLM inference.
    \item We introduce three key architectural innovations: \textbf{1)} a Bidirectional ROM Array (BiROMA) to enhance bit density; \textbf{2)} a Tri-Mode Local Accumulator (TriMLA) to optimize ternary computation and leverage weight sparsity; and \textbf{3)} a Decode-Refresh (DR) eDRAM that buffers part of the KV-cache on-die to reduce memory bandwidth demands. Evaluated in 65nm CMOS, these components collectively enable BitROM to achieve 20.8 TOPS/W energy efficiency (1.5b/4b), a 10$\times$ improvement in bit density over prior digital CiROM designs, and a 43.6\% reduction in memory access during auto-regressive decoding.
    \item We integrate LoRA adapters into BitROM and perform extensive hardware adaptation experiments to identify the optimal configuration for domain transfer. Our results show that only 0.3\% additional quantized LoRA weights are sufficient to enable effective adaptation of BitNet across diverse downstream tasks.
\end{itemize}

\section{Preliminaries}

\subsection{The Era of 1-bit LLMs}

Recent progress in extreme quantization techniques has enabled the emergence of 1-bit LLMs, designed to drastically reduce model size and inference overhead. A representative example is BitNet, which employs quantization-aware training to compress model weights into ternary representations, requiring only $\approx 1.58$ bits per parameter. Notably, BitNet-b1.58~\cite{BitNetb158} maintains perplexity and downstream performance on par with full-precision LLMs while achieving significant reductions in memory footprint, inference latency, and power consumption. The latest variant, BitNet-a4.8~\cite{BitNeta48}, extends this approach by adopting hybrid precision---applying 4b/8b quantization to activations---and introducing pruning to increase weight sparsity, thereby further improving efficiency. However, conventional computing platforms are not optimized to leverage these properties, especially in resource-constrained edge scenarios. This highlights the demand for customized hardware systems specifically designed to accelerate BitNet LLM inference.

\subsection{Large Language Model Inference}

The inference of LLMs follows an auto-regressive generation process, where the model sequentially produces tokens conditioned on prior prompt, as depicted in Fig.~\ref{Fig_Background}(b). This process involves two phases with distinct computational and memory characteristics. \textbf{i)} In the \textbf{prefill phase}, the input prompt is processed in parallel. During this stage, KV pairs are generated for all prompt tokens and stored in the KV-cache for each Transformer Layer. \textbf{ii)} In the \textbf{decode phase}, tokens are generated one at a time. At each step, new KV entries corresponding to the latest token are appended to the existing KV-cache, which is repeatedly accessed to compute the next output. During the decode phase, the size of the KV-cache continuously grows with the sequence length, imposing significant burdens on memory bandwidth and degrading energy efficiency. Consequently, additional efforts are needed to address and optimize this issue.

\subsection{Compute-in-Read-Only-Memory}

CiROM has attracted growing interest in the CiM community due to its ultra-high storage density, offering a path toward eliminating weight update and improving efficiency in edge computing scenarios. Analog CiROM designs prioritize cell density and energy efficiency by exploiting multi-level cells and integrating novel ADC architectures~\cite{Cramming_More_Weight,Compute-MLROM,QLC_CiROM}. However, these designs predominantly target CNN-based image classification tasks and lack robustness for Transformer models requiring higher noise resilience~\cite{3D-METRO,ASiM}. Recent advancements have shifted towards digital CiROM~\cite{DSC-ROM}, achieving a full mapping of ResNet-56 within a 12 mm\textsuperscript{2} silicon footprint using a 65nm CMOS~\cite{DCiROM}. While deterministic digital computing demonstrates the potential to support increasingly complex models, such as Transformers, its chip area remains insufficient to accommodate LLMs due to the sizable adder trees. To mitigate this, a compact CiROM macro that allows more memory cells to share a single adder tree is essential for reducing the area overhead of CiROM when targeting LLMs.

One major limitation of CiROM is that cell weights are fixed during fabrication, resulting in limited flexibility across different tasks. To address this, YOLoC~\cite{YOLoC} introduces ReBrach, incorporating configurable SRAM-based CiM units to enable transfer of pretrained knowledge across vision tasks. Similarly, Hidden-ROM~\cite{Hidden-ROM} integrates configurable SRAM switches to support diverse domains in HNN applications. Hybrid SRAM/ROM CiM architecture~\cite{Hybrid_SRAM_ROM,Hybrid_SRAM_ROM_JSSC} evaluates circuit-level metrics when integrating LoRA adapters~\cite{LoRA} into Transformer models. However, existing software-hardware co-design efforts of CiROM primarily focus on vision domains and CNNs, leaving the exploration of LLM-oriented tasks largely unexplored. In particular, the feasibility of employing LoRA with these ternary-weight BitNet models has yet to be thoroughly investigated, creating uncertainty in the adaptation and optimization strategies and in the selection of suitable hardware configurations for BitNet LLMs.

\section{BitROM Accelerator}

\subsection{System Architecture}

\begin{figure}[htbp]
    \centering
    \includegraphics[width=0.85\columnwidth]{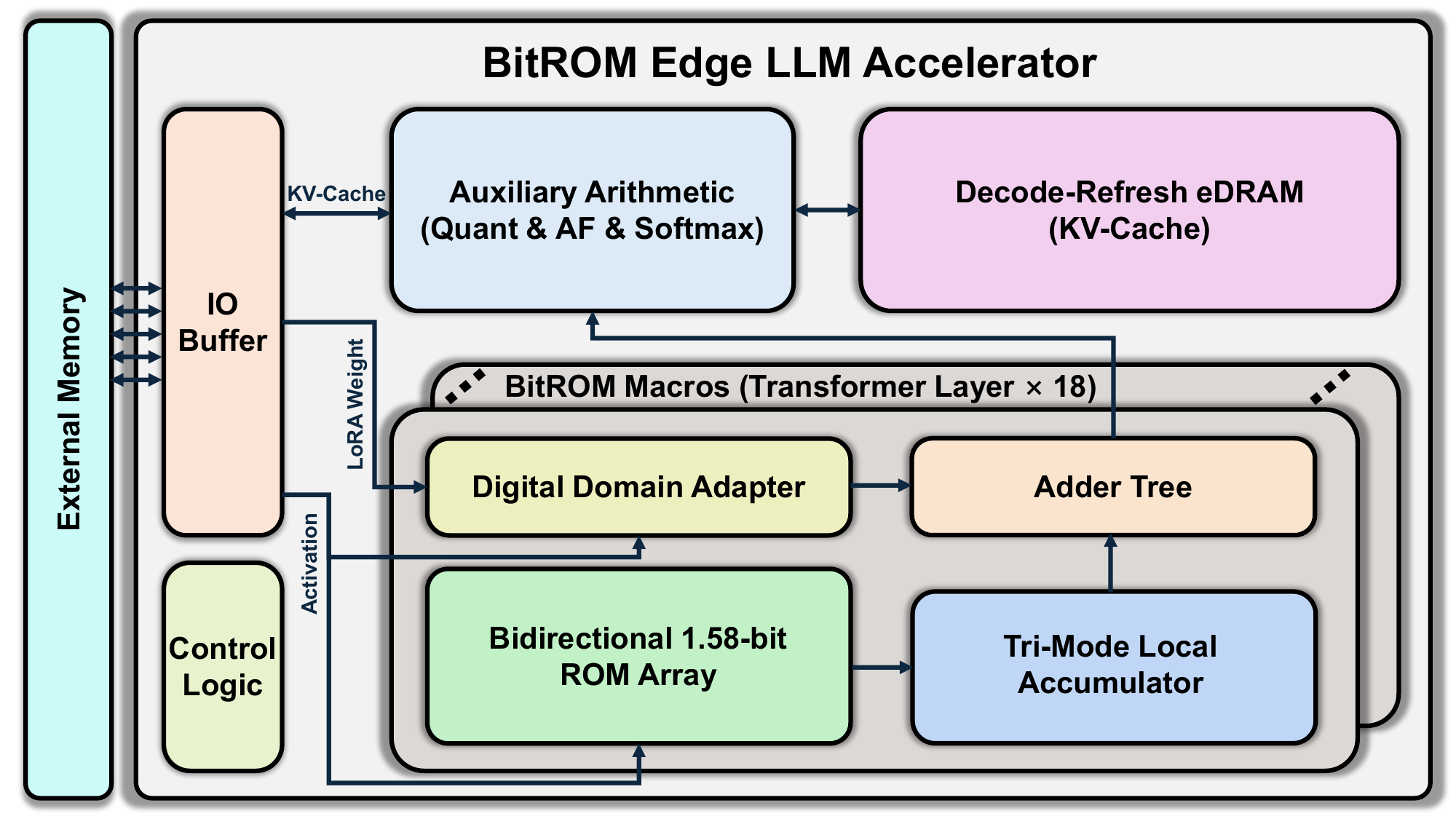}
    \caption{Architecture overview of BitROM accelerator.}
    \label{Fig_Architecture}
\end{figure}

We present the BitROM accelerator for BitNet LLM applications, leveraging a digital CiROM-based approach to fully eliminate weight updates while ensuring error-free, high-quality computation during LLM inference. As illustrated in Fig.~\ref{Fig_Architecture}, the system architecture of BitROM accelerator comprises multiple BitROM macros for weight storage and massive MAC operations in linear projection layers, a DR eDRAM for efficient temporary KV-cache management, an auxiliary arithmetic processor for executing complex floating-point activation functions, quantization, and softmax operations, an I/O buffer to connect with external DRAM, and a dedicated control logic unit that coordinates the entire system.

\subsection{BitROM Macro}

\subsubsection{Motivation and Concept Overview}

\begin{figure}[htbp]
    \centering
    \includegraphics[width=\columnwidth]{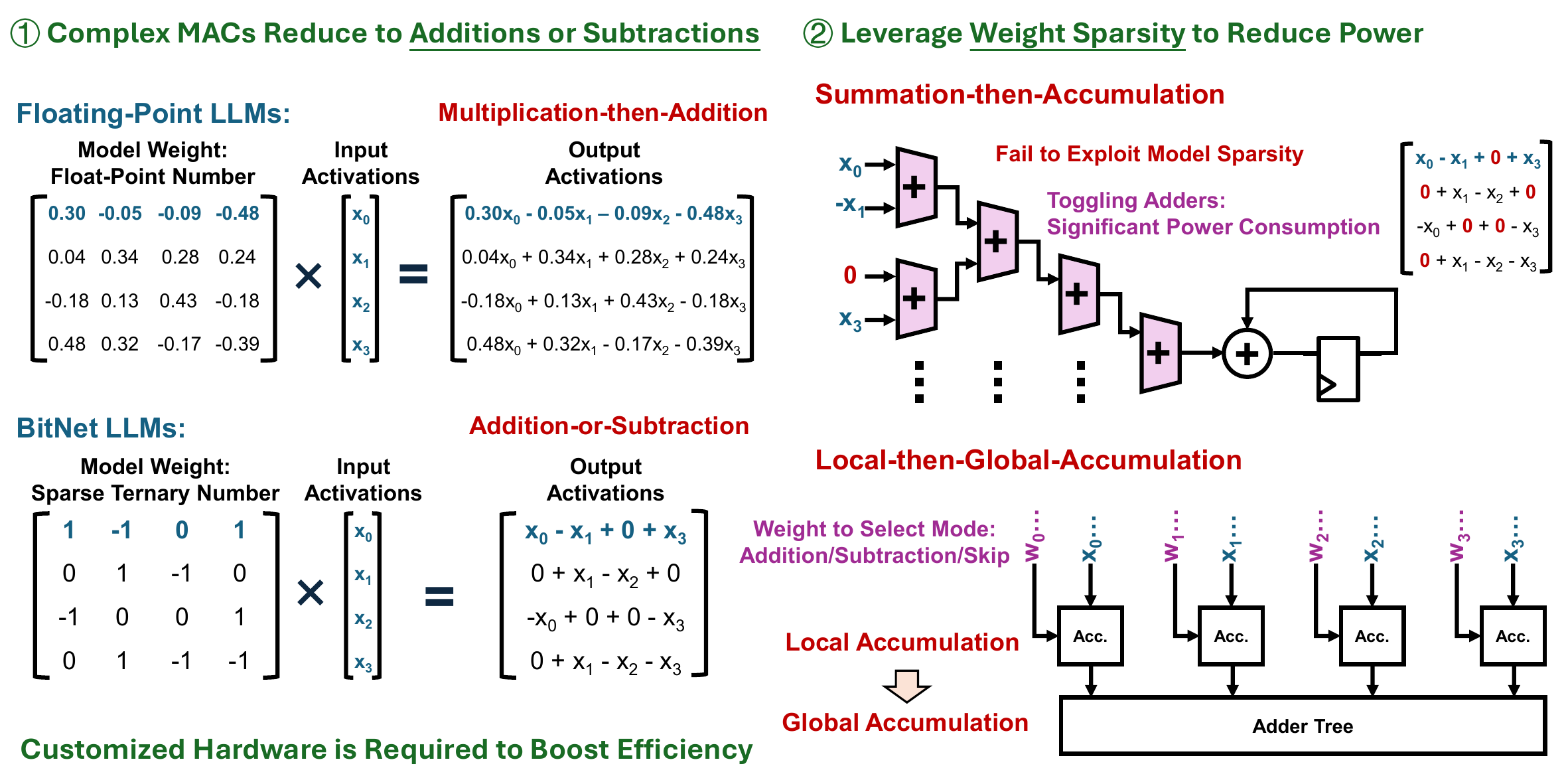}
    \caption{Design motivation and general concept of BitROM.}
    \label{Fig_Motivation}
\end{figure}

BitNet LLMs possess unique features that require careful design optimizations, as illustrated in Fig.~\ref{Fig_Motivation}. Firstly, the ternary-weight representation in BitNet simplifies complex MAC operations into basic additions and subtractions, creating opportunities for circuit-level optimization. Secondly, BitNet models show substantial sparsity, with a significant proportion of zero-valued weights, enabling potential skip operations to lower energy dissipation. Nevertheless, traditional digital CiROM designs adopt a summation-then-accumulation flow using an adder tree, which fails to fully capitalize on the sparsity benefits. Moreover, toggling input bits can propagate through the entire tree, leading to significant energy consumption during each MAC cycle. To address this limitation, we propose a local-then-global-accumulation scheme that is optimized for BitNet LLMs. In this scheme, addition and subtraction operations are first performed sequentially by local accumulators, and only after all channels have been processed does the adder tree execute a single summation. This architecture enables zero-skipping at the local accumulators, leveraging the unique features of BitNet to improve performance and energy efficiency.

\subsubsection{Macro Circuits}

\begin{figure*}[htbp]
\centering
\includegraphics[width=\textwidth]{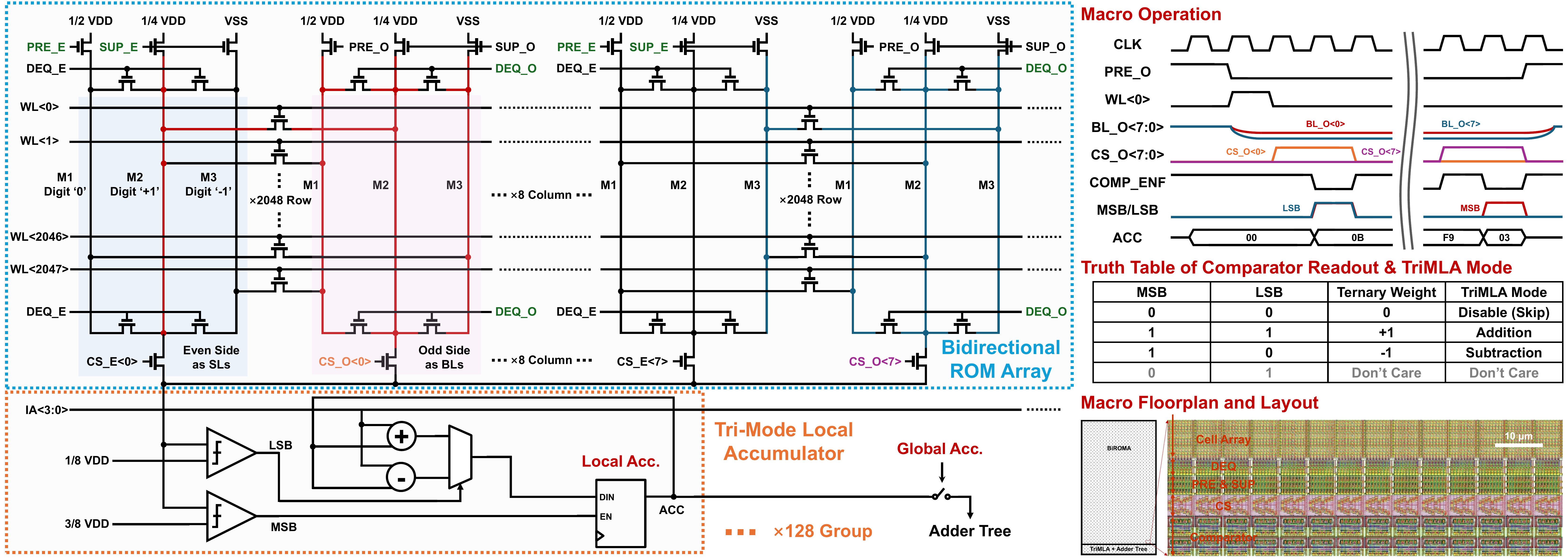}
\caption{BitROM macro implementation and operation. A BiROMA stores two ternary weights per single transistor, while a TriMLA implements the local-then-global accumulation strategy, achieving an energy- and area-efficient digital CiROM design.}
\label{Fig_BiCiROM_Macro}
\end{figure*}

To support high-density digital CiROM storage and implement the local-then-global-accumulation strategy, the BitROM macro comprises two primary components as shown in Fig.~\ref{Fig_BiCiROM_Macro}: BiROMA and TriMLA, where each TriMLA is connected to a group of 8 BiROMA columns. The \textbf{BiROMA} contains 2,048 rows and 1,024 columns, where each ROM cell consists of a single transistor capable of storing two ternary weights, thereby doubling the storage density. The source and drain terminals of each cell are connected to three signal lines representing ternary values '0', '+1', and '-1', corresponding to 1/2 VDD, 1/4 VDD, and VSS, respectively. These signal lines are routed through metal layers M1, M2, and M3 using minimum-pitch design rules to maximize storage density. The signal lines are divided into even (E) and odd (O) sides, each of which can be independently controlled by a voltage supply controll (PRE and SUP), a column selector (CS), and digit equalizers (DEQ). Notably, one side of signal lines can be configured as source lines (SLs) to supply drive signals, while the other functions as bitlines (BLs) for cell readout, thereby enabling bidirectional operation. In \textbf{TriMLA}, ternary weights retrieved from BiROMA are first processed by a pair of comparators with 1/8 VDD and 3/8 VDD reference levels to determine the operational mode. After local accumulation is completed within each TriMLA, the results are further aggregated via an adder tree to perform a one-shot global accumulation.

\subsubsection{Macro Operations}

Fig.~\ref{Fig_BiCiROM_Macro} illustrates the operational flow of the BitROM macro. During weight readout in the BiROMA, one side of the signal lines is configured as SLs, while the other is precharged and equalized to serve as BLs for signal development. Upon wordline (WL) activation, the active cell transistor pulls down the BL signal to capture data from the corresponding SL. Importantly, the even (E) and odd (O) sides of each column are fully symmetric, enabling bidirectional readout and contributing to a compact, efficient digital CiROM implementation. After activating the column selector (CS) switches for each column, the ternary weights are prefetched and used to configure the TriMLA's operating mode according to the truth table shown in Fig.~\ref{Fig_BiCiROM_Macro}. The first comparator output (MSB) determines whether the weight is zero and directly controls the enable (EN) signal of the TriMLA, thereby exploiting sparsity by disabling accumulation when the weight is zero. The second comparator (LSB) selects between addition and subtraction operations. Once all MAC operations for a channel are completed, the outputs of the TriMLAs are aggregated via an adder tree to form the final result. To support both 4-bit and 8-bit BitNet activation formats, TriMLA accepts 4-bit input activations. For 8-bit activation models, bit-serial processing is performed in two cycles with shifting and accumulation. Due to the symmetric distribution of positive and negative weights, the output dynamic range remains limited. Empirical evaluation confirms that an 8-bit output width for TriMLA is sufficient to avoid overflow. We have completed the layout and verification of the entire BitROM macro. The TriMLAs, peripheral logic, and adder tree collectively occupy only 4.8\% of the total area, incurring minimal area overhead. Unlike prior digital CiROM designs where each small group of cells shares a dedicated adder tree~\cite{DCiROM}---resulting in area inefficiency---our BiROMA sequentially processes ternary operations in TriMLA and shares a single adder tree across the entire array. This design achieves a 10$\times$ improvement in area efficiency and enables practical deployment of billion-parameter models.

\subsection{Domain Adaptation}

To enhance the flexibility of BitROM, we employ LoRA adaptation~\cite{LoRA} and design a digital domain adapter to improve performance across various tasks. Within each Transformer block, we integrate a simple 4-input multiplier-and-adder unit to the BitROM macros, enabling efficient MAC computation for LoRA adaptation. Our results demonstrate that employing a LoRA rank of 16 exclusively on the \texttt{Value} and \texttt{Output} projections of the attention block, as well as the \texttt{Down} projection in the MLP block, is sufficient to improve BitNet's performance across multiple tasks. The LoRA weights are quantized to 6 bits, while the activations remain at 8 bits, aligning with the Falcon3 series BitNet configuration~\cite{Falcon3}. Notably, the additional operations account for only 0.7\% of their corresponding projection layers, resulting in negligible power and area overhead.

\begin{figure}[htbp]
    \centering
    \begin{subfigure}{0.49\linewidth}
        \includegraphics[width=\linewidth]{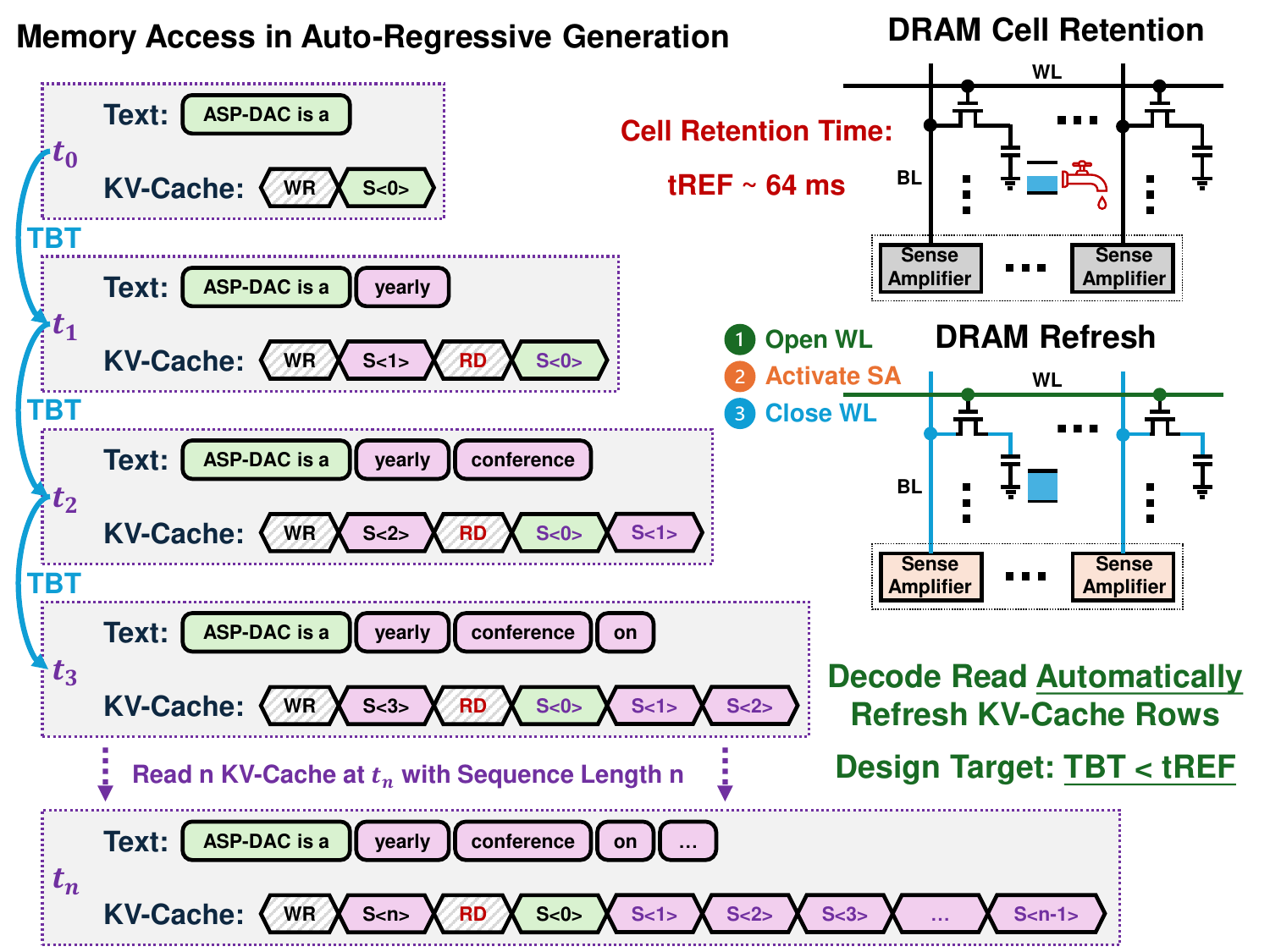}
        \caption{}
        \label{}
    \end{subfigure}
    \hfill
    \begin{subfigure}{0.49\linewidth}
        \includegraphics[width=\linewidth]{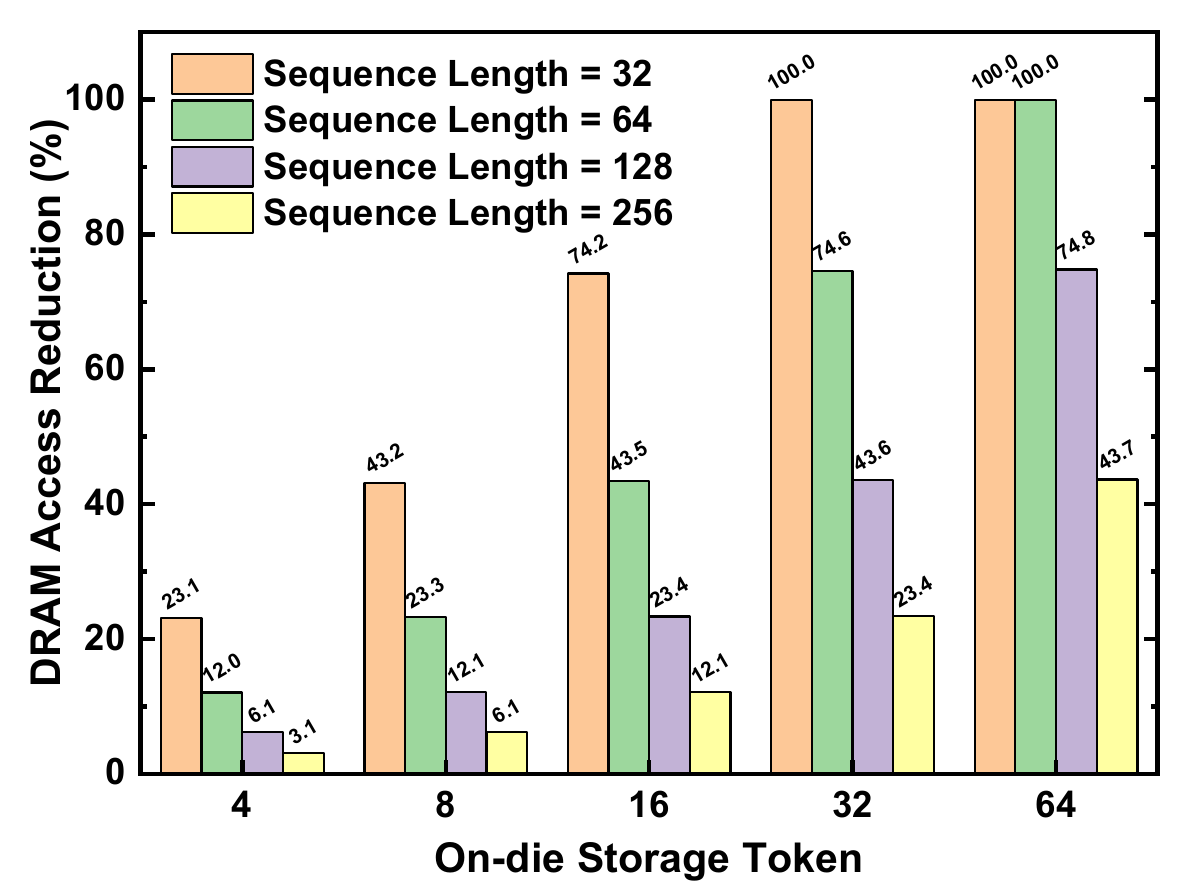}
        \caption{}
        \label{}
    \end{subfigure}
    \caption{DR eDRAM: (a) Analysis of KV-cache behavior and DRAM refresh operations. (b) Reduction in external DRAM accesses using the proposed approach.}
    \label{Fig_Decode-Refresh_eDRAM}
\end{figure}

\begin{table*}[htbp]
\centering
\caption{Comparison of adaptation \textbar{} base results across multiple zero-shot benchmarks. LoRA with a rank of 16 is applied to the \texttt{Value}, \texttt{Output}, and \texttt{Down} projection layers. The results demonstrate sufficient flexibility across various tasks.}
\label{Table_Benchmark}
\begin{tabular}{c|c|c|c|c|c|c|c|c}
\toprule
\multirow{2}{*}{\textbf{Model}} & \textbf{Parameter} & \textbf{WikiText-2} & \textbf{PTB} & \multicolumn{2}{c|}{\textbf{SQuAD}} & \multicolumn{2}{c|}{\textbf{Gigaword}} & \textbf{DROP} \\
\cmidrule(lr){2-2} \cmidrule(lr){3-3} \cmidrule(lr){4-4} \cmidrule(lr){5-6} \cmidrule(lr){7-8} \cmidrule(lr){9-9}
 & \textbf{\%} & \multicolumn{2}{c|}{\textbf{PPL $\downarrow$}} & \textbf{EM $\uparrow$} & \textbf{F1 $\uparrow$} & \textbf{ROUGE-1 $\uparrow$} & \textbf{ROUGE-L $\uparrow$} & \textbf{F1 $\uparrow$} \\
\midrule
\textbf{Falcon3-1B} & 0.30 & 20.45 \textbar{} 20.79 & 52.58 \textbar{} 53.82 & 4.94 \textbar{} 2.49 & 23.09 \textbar{} 19.30 & 25.84 \textbar{} 23.67 & 23.25 \textbar{} 21.09 & 9.87 \textbar{} 8.37 \\
\textbf{Falcon3-3B} & 0.25 & 16.57 \textbar{} 16.99 & 40.12 \textbar{} 41.99 & 1.03 \textbar{} 0.02 & 15.53 \textbar{} 14.43 & 21.38 \textbar{} 19.43 & 19.29 \textbar{} 17.33 & 4.31 \textbar{} 3.90 \\
\textbf{Falcon3-7B} & 0.22 & 14.84 \textbar{} 15.10 & 34.78 \textbar{} 36.25 & 12.13 \textbar{} 3.80 & 40.84 \textbar{} 32.52 & 23.33 \textbar{} 20.86 & 21.15 \textbar{} 18.88 & 15.09 \textbar{} 10.76 \\
\textbf{Falcon3-10B} & 0.23 & 17.86 \textbar{} 18.28 & 41.88 \textbar{} 43.40 & 9.64 \textbar{} 1.98 & 35.70 \textbar{} 28.35 & 30.68 \textbar{} 28.72 & 28.21 \textbar{} 26.25 & 12.10 \textbar{} 7.95 \\
\bottomrule
\end{tabular}
\end{table*}

\section{Decoding-Aware KV-cache Management}

Distinct from conventional CNN or Transformer-based models, the auto-regressive nature of LLMs leads to a considerably more complex inference process. The key difference lies in the decode phase, during which the activations of previous \texttt{Key} and \texttt{Value} projections are reused at each subsequent decoding step in the auto-regressive generation process. As decoding advances, the KV-cache continuously expands, creating substantial pressure on memory bandwidth and constraining overall system performance. To elucidate the decoding process and identify potential optimization opportunities, we decompose an LLM inference step by step, as depicted in Fig.~\ref{Fig_Decode-Refresh_eDRAM}(a). This analysis reveals two unique properties of the KV-cache: \textbf{i)} At each decoding step, a single memory write operation occurs, whereas the number of memory read operations increases linearly with the number of decoding steps. \textbf{ii)} Once the KV-cache of a token is written, it is subsequently accessed at every decoding step thereafter, leading to early tokens being read more frequently. For instance, the prompt tokens “\textit{ASP-DAC is a}” (S$<$0$>$) are read $n$ times during the generation of a sequence of length $n$, and the first token “\textit{yearly}” (S$<$1$>$) is read $n-1$ times after decoding step $t_2$. Recognizing that not all tokens have an equivalent impact on memory bandwidth, we propose that buffering a portion of the KV-cache for early tokens on-die can achieve notable efficiency gains compared to relying solely on external memory.

\begin{table}[htbp]
\centering
\caption{Ablation study of adapters across different layers, showing that the \texttt{Value}, \texttt{Output}, and \texttt{Down} projection layers best balance additional operations and performance.}
\label{Table_Ablation_Layer}
\begin{tabular}{c|c|c|c|c|c|c|c|c|c}
\toprule
\textbf{Q} & \textbf{K} & \textbf{V} & \textbf{O} & \textbf{G} & \textbf{U} & \textbf{D} & \textbf{Parameter} & \textbf{EM} & \textbf{F1} \\
\midrule
$\checkmark$ & $\checkmark$ & $\times$ & $\times$ & $\checkmark$ & $\checkmark$ & $\times$ & 0.37\% & 3.87 & 33.05 \\
$\times$ & $\times$ & $\times$ & $\times$ & $\times$ & $\times$ & $\checkmark$ & 0.16\% & 8.04 & 37.97 \\
$\times$ & $\times$ & $\times$ & $\checkmark$ & $\times$ & $\times$ & $\checkmark$ & 0.19\% & 11.23 & 40.37 \\
$\times$ & $\times$ & $\checkmark$ & $\checkmark$ & $\times$ & $\times$ & $\checkmark$ & 0.22\% & 12.13 & 40.84 \\
$\checkmark$ & $\checkmark$ & $\checkmark$ & $\checkmark$ & $\checkmark$ & $\checkmark$ & $\checkmark$ & 0.59\% & 13.61 & 41.67 \\
\bottomrule
\end{tabular}
\end{table}

To enable efficient on-die KV-cache storage, the memory must offer both high speed and high density, positioning eDRAM as the most practical option available today. However, as DRAM is volatile, data loss due to cell retention issues necessitates refresh management, which complicates system design and reduces power efficiency~\cite{Refresh_Trade-offs}. On the other hand, the DRAM refresh process is inherently simple: \textbf{i)} opening the WL at the target address, \textbf{ii)} activating the associated sense amplifier, and \textbf{iii)} closing the WL to complete the refresh, as depicted in Fig.~\ref{Fig_Decode-Refresh_eDRAM}(a). Importantly, whenever an address is accessed for reading, its contents are automatically refreshed. Considering the nature of KV-cache, once the KV-cache of a token is stored in eDRAM, it is accessed in every subsequent decoding step, ensuring automatic refresh and eliminating the need for explicit refresh management in LLM inference. The only requirement is to ensure that the Token-Between-Token (TBT) timing is shorter than the DRAM cell retention time (tREF), typically 64 ms~\cite{JEDEC_DDR5}, which aligns well with general TBT values in LLM inference. Given that the BitROM is designed for edge applications and short-sequence tasks (such as instruction execution and question answering), partial on-die KV-cache storage is both feasible and beneficial. We therefore leverage a straightforward DR eDRAM in the BitROM to store the early tokens of a sequence, thereby achieving notable efficiency improvements. In our study, we consider sequence lengths ranging from 32 to 256 in typical edge applications and examine the impact of moving 4 to 64 early tokens to on-die storage. Fig.~\ref{Fig_Decode-Refresh_eDRAM}(b) depicts the resulting reduction rates in external DRAM read operations. The findings reveal that relocating only 1/4 of the early tokens to on-die storage can reduce the DRAM access rate by nearly half. For example, at a sequence length of 128, BitROM achieves a 43.6\% reduction in external DRAM access by storing 32 early tokens on-die.

\section{Experiments}

\subsection{Adaptation Experiment}

To assess the adaptability of BitROM across a range of tasks, we evaluate perplexity on WikiText-2 and PTB for language modeling; EM and F1 scores on SQuAD for question answering; ROUGE-1 and ROUGE-L scores on Gigaword for text summarization; and the F1 score on DROP for paragraph comprehension. These results are presented in Table~\ref{Table_Benchmark}. In our implementation, LoRA is applied exclusively to the \texttt{Value} and \texttt{Output} projections within the attention blocks and the \texttt{Down} projection in the MLP blocks. The LoRA weights are quantized to 6 bits, while the activations remain at 8 bits. We set the LoRA rank to 16, introducing only 16 additional neurons per layer---an amount negligible compared to their channel dimensions ranging from 2,048 to 8,192. Following LoRA adaptation, the models consistently outperform the baseline models, demonstrating BitROM's strong domain transfer capability with minimal overhead. Interestingly, we observe that the performance improvement scales with the base model size; for instance, Falcon3-7B exhibits EM and F1 score gains of 8.33 and 8.32, respectively, highlighting the flexibility of the BitROM.

\begin{table*}[htbp]
\centering
\caption{Comparison of BitROM performance against state-of-the-art hardware accelerators.}
\label{Table_Comparison}
\begin{tabular}{c|ccccccc}
\toprule
 & \textbf{ISSCC '25~\cite{Slim-Llama}} & \textbf{JSSC '23~\cite{Cramming_More_Weight}} & \textbf{ESSCIRC '23~\cite{Compute-MLROM}} & \textbf{ASSCC '24~\cite{QLC_CiROM}} & \textbf{CICC '24~\cite{Hybrid_SRAM_ROM}} & \textbf{ASPDAC '25~\cite{DCiROM}} & \textbf{This Work} \\
\midrule
\textbf{Technology}   & 28 nm & 65 nm & 65 nm & 28 nm & 28 nm & 65 nm & 65 nm \\
\textbf{Domain} & Digital & Analog & Analog & Analog & Analog & Digital & Digital \\
\textbf{Voltage (V)} & 0.65 & 0.7/1.2 & 1.1 & 0.6 & 0.7/1.1 & 0.6/1.2 & 0.6/1.2 \\
\textbf{Model Type} & 1.58b/4b & 8b/8b & 2b/1b & 8b/8b & 8b/8b & 4b/4b & 1.58b/4b\\
\textbf{Bit/Cell} & - & 2 & 2 & 4 & 2 & 1 & 1.58$\times$2 \\
\textbf{Eff. (TOPS/W)} & 255.9 & 4.33/1.24 & 1,324.26 & 8.49 & 42.0/20.3 & 38.0/9.0 & 20.8/5.2 \\
\textbf{\textsuperscript{*}Norm. Eff.} & 47.5 & 4.33/1.24 & 1,324.26 & 1.58 & 7.8/3.8 & 38.0/9.0 & 20.8/5.2 \\
\textbf{Bit Density} & - & 3,984 Kb/mm\textsuperscript{2} & 375 Kb/mm\textsuperscript{2} & 19,660 Kb/mm\textsuperscript{2} & 8,928 Kb/mm\textsuperscript{2} & 487 Kb/mm\textsuperscript{2} & 4,967 Kb/mm\textsuperscript{2} \\
\textbf{\textsuperscript{*}Norm. Den.} & - & 3,984 Kb/mm\textsuperscript{2} & 375 Kb/mm\textsuperscript{2} & 3,648 Kb/mm\textsuperscript{2} & 1,657 Kb/mm\textsuperscript{2} & 487 Kb/mm\textsuperscript{2} & 4,967 Kb/mm\textsuperscript{2} \\
\textbf{KV Optm.} & $\times$ & $\times$ & $\times$ & $\times$ & $\times$ & $\times$ & -43.6\% \\
\textbf{Update-Free} & $\times$ & $\checkmark$ & $\checkmark$ & $\checkmark$ & $\checkmark$ & $\checkmark$ & $\checkmark$ \\
\bottomrule
\end{tabular}
\makebox[\linewidth][l]{\footnotesize\textsuperscript{*} Energy efficiency and bit density values are normalized to a 65nm CMOS process based on spatial scaling ratios.}
\end{table*}

We further perform ablation studies on SQuAD by inserting rank-16 adapters into different layer combinations of the Falcon3-7B model, as shown in Table~\ref{Table_Ablation_Layer}. The findings reveal that the effectiveness of adaptation strongly depends on the chosen layers. For example, adding adapters to the \texttt{Query} and \texttt{Key} projections in the attention blocks or to the \texttt{Gate} and \texttt{Up} projections in the MLP blocks results in performance nearly indistinguishable from the base model, despite the additional parameters comprising 0.37\%. In contrast, our configuration---which involves placing adapters at the \texttt{Value}, \texttt{Output}, and \texttt{Down} projections—achieves performance nearly equivalent to that of full adaptation while requiring only 0.22\% additional parameters, underscoring the efficiency of our approach.

\begin{figure}[htbp]
    \centering
    \begin{subfigure}{0.49\linewidth}
        \includegraphics[width=\linewidth]{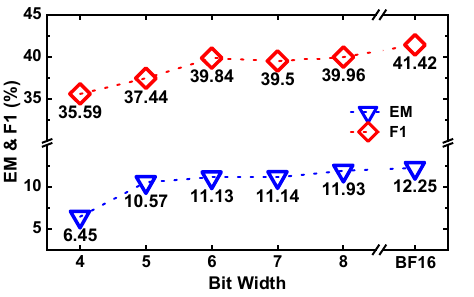}
        \caption{}
        \label{}
    \end{subfigure}
    \hfill
    \begin{subfigure}{0.49\linewidth}
        \includegraphics[width=\linewidth]{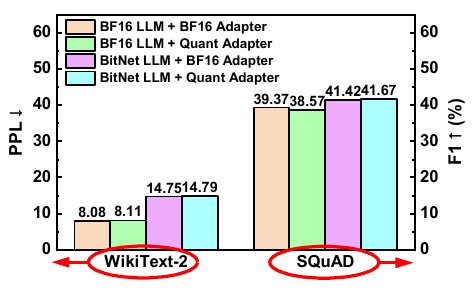}
        \caption{}
        \label{}
    \end{subfigure}
    \caption{Ablation studies: (a) Dependence of adaptation performance on varying quantization bit widths; (b) Comparison of BitNet with full-precision LLM.}
    \label{Fig_Ablation}
\end{figure}

We also evaluate the impact of model quantization on performance for SQuAD using Falcon3-7B, as shown in Fig.~\ref{Fig_Ablation}(a). With activations fixed at 8 bits to match the model's activation bit width, we examine how the quantization bit width of LoRA weights affects downstream performance. The results demonstrate that 6-bit quantization for LoRA weights is sufficient to maintain high EM and F1 scores, ensuring minimal overhead and robust performance of the adapters in the BitROM. In addition, we investigate the impact of quantization on both the adapter and LLM models, with findings illustrated in Fig.~\ref{Fig_Ablation}(b). We observe that quantization of the adapter consistently has negligible impact on performance in both full-precision and BitNet LLMs. Notably, while BitNet LLMs exhibit higher perplexity compared to their full-precision counterparts, they unexpectedly achieve better task performance---likely due to reduced overfitting resulting from the extreme quantization.

\subsection{System Performance}

We implemented the BitROM accelerator using the TSMC 65nm standard digital CMOS process. To assess the performance of the DR eDRAM integration, we adopt the eDRAM design reported in~\cite{GC-eDRAM} for performance analysis. Targeting the Falcon3-1B model~\cite{Falcon3}, which features 18 Transformer layers and 4 key-value heads via grouped-query attention, the BitROM accelerator is architected with 6 independent macro partitions. Each partition maps to 3 Transformer layers and supports up to 6 input batches. Batch-level computations are distributed across a 6-stage pipeline, allowing all partitions to operate in parallel and maintain full macro utilization. For an edge inference scenario with a sequence length of 128 and buffering of 32 early KV-cache tokens, we allocate 13.5 MB of DR eDRAM to meet the memory requirements.

We evaluate the performance of the BitROM accelerator through post-layout simulations and compare it with several related designs, as summarized in Table~\ref{Table_Comparison}. BitROM achieves a high energy efficiency of 20.8 TOPS/W and an ultra-compact bit density of 4,967 kB/mm\textsuperscript{2}, demonstrating strong competitiveness against state-of-the-art accelerators. Unlike analog CiROMs, BitROM avoids computing inaccuracies and supports robust inference. While \cite{Compute-MLROM} and \cite{DCiROM} report higher energy efficiency, BitROM delivers over 10$\times$ higher bit density, highlighting its compatibility with LLM-scale models. Furthermore, while digital accelerator optimized for BitNet~\cite{Slim-Llama} shows higher energy efficiency, BitROM eliminates weight update overhead through its CiROM architecture, leading to improved system-level energy efficiency. Importantly, BitROM incorporates on-die KV-cache management in the decode phase, reducing external memory access by up to 43.6\% during auto-regressive generation---paving the way for efficient edge deployment of LLMs. Finally, we assess BitROM's scalability on advanced technology nodes. When implemented in a 14nm CMOS for the Falcon3-1B model, BitROM and its on-die DR eDRAM occupy only 16.71 and 10.24 cm\textsuperscript{2}, respectively, making ultra-low-power LLM deployment on edge devices feasible.

\section{Conclusion}

In this paper, we introduced BitROM, a CiROM-based accelerator tailored for BitNet LLMs for edge applications. BitROM integrates a high-density BiROMA and a TriMLA architecture to enable efficient and compact ternary computation. To address the memory bottleneck introduced by KV-cache during auto-regressive decoding, BitROM incorporates a DR eDRAM, which effectively reduces external memory traffic by caching projection of early tokens on-die. Additionally, we adopt a LoRA-based domain adaptation strategy, enabling task flexibility with minimal hardware overhead. BitROM achieves 20.8 TOPS/W energy efficiency and a 10$\times$ improvement in bit density over prior digital CiROM designs, positioning it as a promising step toward enabling billion-parameter BitNet deployment for energy-efficient LLM inference.

\section*{Acknowledgments}
This research was supported in part by the JST CREST JPMJCR21D2, JSPS Kakenhi 23H00467, KIOXIA Encouragement Research Grant, JST Doctoral Program Student Support Project, Futaba Foundation, Asahi Glass Foundation, and Telecommunications Advancement Foundation.

\bibliographystyle{IEEEtran}
\bibliography{reference}

\end{document}